\documentclass[useAMS,usenatbib,usegraphicx]{mn2e}
\usepackage{epsf}

\title[The image separation distribution of strong lenses]
{The image separation distribution of strong lenses: \\
Halo versus subhalo populations}
\author[M. Oguri]
{Masamune Oguri$^{1, 2}$\thanks{E-mail:oguri@astro.princeton.edu} \\
$^1$Princeton University Observatory, Peyton Hall,
Princeton, NJ 08544, USA.\\
$^2$Department of Physics, University of Tokyo, Hongo
7-3-1, Bunkyo-ku, Tokyo 113-0033, Japan.}
\begin{document}

\date{\today}

\voffset- .65in

\pagerange{\pageref{firstpage}--\pageref{lastpage}} \pubyear{2004}

\maketitle

\label{firstpage}

\begin{abstract}
We present a halo model prediction of the image separation distribution 
of strong lenses. Our model takes account of the subhalo population, which
has been ignored in previous studies, as well as the conventional halo 
population. Halos and subhalos are linked to central and satellite galaxies 
by adopting an universal scaling relation between masses of (sub-)halos and 
luminosities  of galaxies. Our model predicts that $10\%-20\%$ of lenses 
should be caused by the subhalo population. The fraction of lensing by 
satellite galaxies (subhalos) peaks at $\sim 1''$ and decreases rapidly 
with increasing image separations. We compute fractions of lenses which
lie in groups and clusters, and find them to be $\sim 14\%$ and $\sim
4\%$, respectively: Nearly half of such lenses are expected to be
produced by satellite galaxies, rather than central parts of halos.
We also study mass distributions of lensing halos and find that even
at image separations of $\sim 3''$ the deviation of lens mass
distributions from isothermal profiles is large: At or beyond $\sim 3''$
image separations are enhanced significantly by surrounding halos. Our
model prediction agrees reasonably well with observed image separation
distributions from galaxy to cluster scales. 
\end{abstract}


\section{Introduction}

Image separations between multiple images play an important role
in the statistics of lensed quasars: The image separation is mainly
determined by the potential wall depth of the lens object, thus the
image separation distribution reflects hierarchical structure in the
universe. This fact suggests that lensed quasars may categorized roughly
into two populations; small-separation lens ($\theta\sim 1''$) that are
produced by a single galaxy, and a large-separation lens ($\theta\ga
10''$) that are caused by clusters of galaxies. Thus far about 80 lensed
quasars are known: Most are small-separation lenses and only one lens,
SDSS J1004+4112 \citep{inada03,oguri04a}, has the image separation larger
than $10''$. The current largest homogeneous lens survey, the Cosmic
Lens All Sky Survey \citep[CLASS;][]{myers03,browne03}, contains small
separation only, thus it may not be suitable for studying the image
separation distribution from galaxy to cluster scales. However, ongoing
lens surveys such as that in the Sloan Digital Sky Survey (SDSS) are
expected to unveil the full distribution, which will be extremely useful
for understanding the assembly of structures.  

In theoretical sides, there have been many attempts to calculate the
full image separation distribution. Historically, the image 
separation distribution is computed from either the galaxy luminosity
function \citep{turner84} or the mass function of dark halos
\citep{narayan88}. While the former approach accounts for the
observed lensing probability and its distribution at $\theta<3''$ fairly
well, it has difficulty in explaining the existence of the
large-separation lens for which dark matter dominates the lens
potential. Therefore in this paper we take the latter approach.
 
However, modeling the image separation distribution on the basis of dark
halos is not simple. It has been found that none of models that consider
only one population for lensing halos are consistent with current
observations. Correct models need to include a characteristic mass around
$M_{\rm cool}\sim 10^{13}M_\odot$, where the density distribution inside
dark halos changes 
\citep{porciani00,kochanek01,keeton01,sarbu01,li02,li03,oguri02b,ma03,huterer04,kuhlen04,chen04,zhang04,chen05,zhang05}: 
Large mass halos $M>M_{\rm cool}$ (correspond to groups or clusters) 
have cooling time longer than the Hubble time, thus the internal
structure of dark halos is not strongly affected by baryon cooling and
is well described by that of dark matter \citep{navarro97}. On the other
hand, in small mass halos $M<M_{\rm cool}$ (correspond to galaxies)
the baryon cooling is very efficient and the mass distribution becomes
strongly centrally concentrated, which is well approximated by the
Singular Isothermal Sphere (SIS). Although this two-component model is
successful in explaining the image separation distribution to some
extent, clearly it is not sufficient. First of all, this model considers
only galaxies that lie at the center of isolated dark halos, and thus
does not take galaxies in groups and clusters into account. Put another
way, this model neglects substructures (subhalos) which should
corresponds to satellite galaxies. Actually, it
has been speculated that lens systems in such dense environments is
quite common \citep{keeton00}. In addition, sometimes lens systems  
are significantly affected by the mass distribution outside the lens
object. A good example is Q0957+561 \citep{walsh79}; its lens object is
a galaxy in a cluster, and the image separation is much larger than 
expected from the luminosity of the lens galaxy because the image
separation is boosted by the cluster potential. Indeed, it has been
shown theoretically that at intermediate-separation regime
($\theta=3''-7''$) lens galaxies in dense environments are quite
common \citep{oguri05}. Therefore the correct model needs to account
for such external mass. 

To construct a model that is based on dark halos and subhalos, we
have to model the relation between mass of dark halos (subhalos)
and luminosities of galaxies inside them. There are several models to
link them: Conditional luminosity function (CLF) which is defined by the
luminosity function in a halo of given mass
\citep{yang03,vandenbosch05}, universal mass-luminosity relation between
halo/subhalo mass and hosted galaxy luminosity \citep{vale04}, and the
mixture of these two models \citep{cooray05a,cooray05b}. There models
are calibrated to match observed galaxy luminosity functions, clustering of
galaxies as a function of galaxy luminosities, and the luminosity
function of galaxies in groups or clusters. 

In this paper, we study the image separation distribution with a
particular emphasis 
on the different contributions from central and satellite galaxies. For
central galaxies, we consider baryon cooling in a dark matter halo as
done by \citet{kochanek01} and \citet{keeton01}. Satellite galaxies,
which have been neglected in previous studies, are linked to dark halo
substructures. We basically take the approach proposed by \citet{vale04}
to relate mass of dark halos and subhalos with luminosities of
galaxies, but we modify it to account for the  relatively small
abundance of satellite galaxies in galactic halos
\citep[see][]{cooray05b}. For satellite galaxies, the external mass from
the host halo, which is shown to be important at large-image
separations, is taken into account. In addition to the difference
between central and satellite galaxies, we also pay particular attention
to how lens objects make the transition from SIS to NFW as halo masses
increase.
 
This paper is organized as follows. In \S \ref{sec:halo}, we introduce
models of dark halos and subhalos that are used to compute lensing
probabilities. Section \ref{sec:link} are devoted to model the relation
between (sub-)halo masses and galaxy luminosities. We compute the image
separation distributions in \S \ref{sec:p}, and show the results in
\S \ref{sec:sepdist}. 
Finally, we summarize and discuss our results in \S \ref{sec:sum}.
Throughout the paper we assume a
concordance cosmology with the mass density $\Omega_M=0.3$, the 
cosmological constant $\Omega_\Lambda=0.7$, the dimensionless Hubble
parameter $h=0.7$, and the normalization of the mass fluctuation
$\sigma_8=0.9$. 

\section{Modeling Dark Halos and Subhalos}\label{sec:halo}

\subsection{Dark Halos}\label{sec:dh}

For the density profile of original dark halos, we adopt the spherical
NFW profile \citep{navarro97}: 
\begin{equation}
 \rho(r)=\frac{\rho_s}{(r/r_{\rm s})(1+r/r_{\rm s})^2}.
\label{nfw}
\end{equation}
The characteristic density $\rho_{\rm s}$ is computed 
from the nonlinear overdensity $\Delta_{\rm vir}$. The scale radius
$r_{\rm s}$ is related to the concentration parameter $c=r_{\rm
vir}/r_{\rm s}$, where $r_{\rm vir}$ is the virial radius.  
The median concentration parameter depends on the mass $M$ and redshift $z$ of
the dark halo: We adopt the fitting form derived by \citet{bullock01}:
\begin{equation}
 \bar{c}=\frac{10}{1+z}\left(\frac{M}{M_*}\right)^{-0.13},
\label{bullock}
\end{equation}
where $M_*$ is determined by solving $\sigma_M = \delta_{\rm c}$ at $z=0$
($\sigma_M$ is the standard deviation of linear density field smoothed with a
top-hat filter enclosing mass $M$, and $\delta_{\rm c}\approx 1.68$ is the
threshold linear overdensity above which the region collapses). 
We also include the scatter in the concentration parameter, which is
well described by a  log-normal distribution:
\begin{equation}
 p(c)=\frac{1}{\sqrt{2\pi}\sigma_{\ln c}c}
\exp\left[-\frac{(\ln c-\ln \bar{c})^2}{2\sigma_{\ln c}^2}\right],
\label{p_c}
\end{equation}
where the standard deviation is $\sigma_{\ln c}=0.3$.

In this paper, we consider the modification of mass distribution in dark
halos due to baryon cooling. Traditionally, the response of
dark matter to baryon infall has been calculated by the model of
adiabatic contraction \citep{blumenthal86}. The model assumes the
spherical symmetry and computes the response by imposing the conservation
of angular momentum between before and after baryon infall. Recently,
\citet{gnedin04} studies the validity of the model using high-resolution
numerical simulations, and found that the model can be improved by
taking the eccentric orbits of particles into account. They also presented
a series of analytic fitting functions which make it much easier to
implement the modification. We adopt this modified adiabatic compression
model to derive the total mass distribution. For the final mass distribution
of the cooled baryonic component, we assume \citet{hernquist90} profile:
\begin{equation}
 \rho_{\rm b}(r)=\frac{M_{\rm b}}{2\pi}\frac{1}{(r/r_{\rm b})(r_{\rm b}+r)^3},
\label{hernquist}
\end{equation}
which is known to describe the stellar density profile of elliptical
galaxies well. Note that to compute the final mass distribution we need
to specify the cooled mass $M_{\rm b}$ (or equivalently mass fraction
$f_{\rm b}\equiv M_{\rm b}/M$) and the core radius $r_{\rm b}$
(or equivalently the ratio of the core radius to the scale radius
$x_{\rm b}\equiv r_{\rm b}/r_{\rm s}$), in addition to halo parameters
such as $c$ and $M$. 

For the mass function of dark halos, we use the form proposed by
\citet{sheth99}:
\begin{eqnarray}
\frac{dn}{dM}&=&\frac{\rho(z)}{M^2}\frac{d\ln\sigma_M^{-1}}{d\ln M}\\
&&\times A\sqrt{\frac{2a}{\pi}}\left[1+\left(\frac{\sigma_M^2}{a\delta_c^2}\right)^p\right]\frac{\delta_c}{\sigma_M}\exp\left[-\frac{a\delta_c^2}{2\sigma_M^2}\right],
\end{eqnarray}
where $\rho(z)$ is the mean matter density at redshift $z$, $A=0.322$,
$a=0.707$, and $p=0.3$. The mass function is the modification of
\citet{press74} to account for ellipsoidal nature of gravitational
collapse in cosmological situations, and is known to reproduce the mass
distribution and its redshift evolution in numerical simulations
\citep[e.g.,][]{reed03}. 

\subsection{Subhalos}\label{sec:sub}

We consider subhalos which lie in a dark halo with mass $M$. The mass of
the subhalo is denoted by $m_s$ and the distance from the halo
center by $r$. We always adopt an SIS for the density distribution of
galaxies associated with subhalos  for simplicity. Indeed it has
been known that the central part of the density profile of total (dark
matter plus cooled baryon) matter in a dark halo after baryon infall is
quite close to an SIS. For the mass and spatial distributions of
subhalos, we use a model constructed by \citet{oguri04c}. 
In the paper, they showed that the analytic model reproduces well the
results of high-resolution numerical simulations.
In reality, we use the following fitting formulae of mass and spatial
distributions to speed up the computation.

\begin{figure}
\begin{center}
 \includegraphics[width=1.0\hsize]{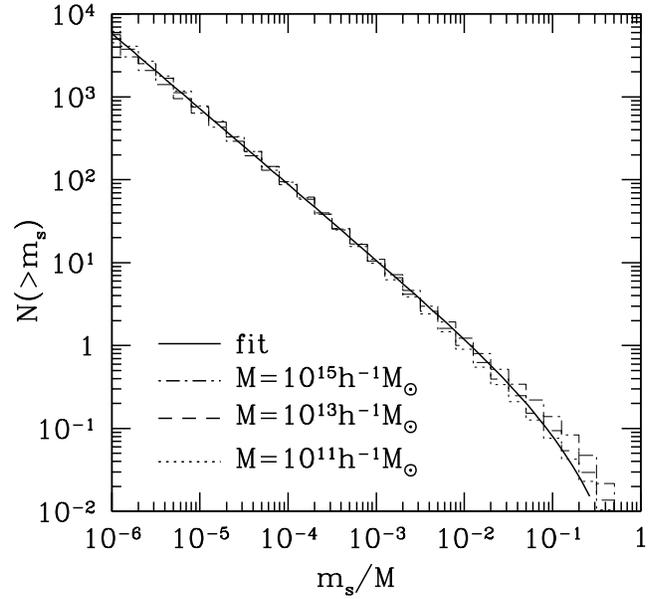}
\end{center}
\caption{The comparison of the cumulative number distributions of
 subhalos derived from an analytic model of \citet{oguri04c} with
 the fitting function (obtained by integrating equation (\ref{mf_sub})
 over $m_s$). The former is shown by histograms, and the latter is
 plotted by a solid line. We plot the distributions of the analytic
 model for three different masses of the host halos. Note that the
 cumulative mass distribution derived from the fitting function does
 not depend on the mass of the host halo when it is plotted as a
 function of $m_s/M$. \label{fig:cum_mass}}
\end{figure}

\begin{figure*}
\begin{center}
 \includegraphics[width=0.8\hsize]{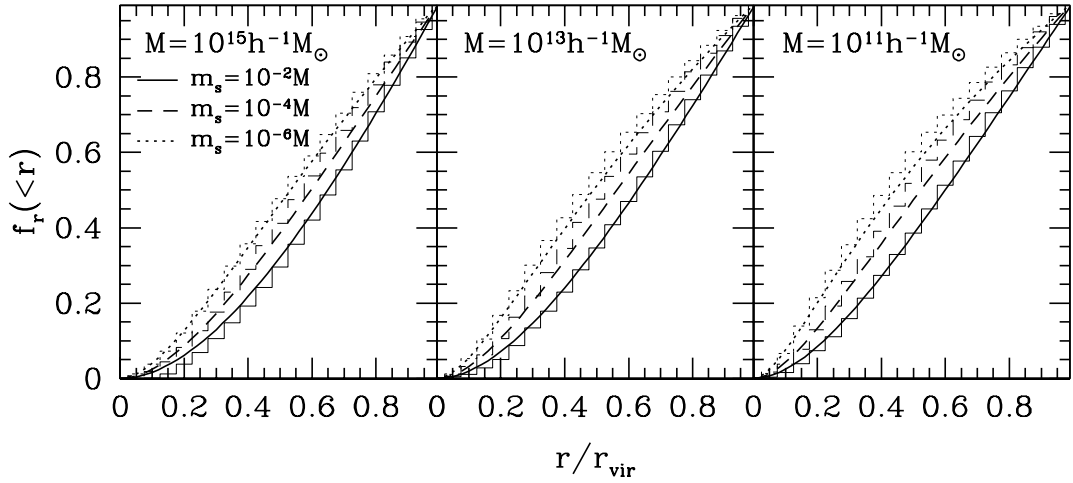}
\end{center}
\caption{The cumulative radial distributions of subhalos in an
 analytic model \citep{oguri04c} are compared with those of an NFW fit
 with concentration parameters given by equation (\ref{c_sub}). The
 former distributions are plotted with histograms, while the latter are
 shown by curves. From left to right panels, we change the mass of the
 host halo. Different lines indicate different masses of subhalos.
\label{fig:cum_rad}}
\end{figure*}

The fitting function of the number distribution of substructures was given by
\citet{vale04}: 
\begin{equation}
\frac{dN}{dm_s}=\frac{0.18}{\gamma\Gamma(2-\beta)}\left(\frac{m_s}{\gamma M}\right)^{-\beta}\exp\left(-\frac{m_s}{\gamma M}\right)\frac{1}{\gamma M},
\label{mf_sub}
\end{equation}
where $\beta=1.91$ and $\gamma=0.39$. The mass function has a power-law
form at $m_s\ll M$, and has a sharp cut-off at $m_s\sim M$. In addition, 
we assume that $dN/dm_s=0$ at $m_s>M$. We compare this fitting function
with the result of \citet{oguri04c} in Figure \ref{fig:cum_rad},
confirming the good accuracy of the fitting function. 

It has been shown that the radial number density distributions of
subhalos are less centrally concentrated relative to the dark
matter particle components. Moreover, they depend on masses of
subhalos: More massive subhalos are preferentially located in 
the external regions of their host halos
\citep{delucia04,oguri04c,gao04}. We find that the spatial
distribution of subhalos is described by an NFW profile
(eq. [\ref{nfw}]) with different concentration parameter (which depend
on $m_s$) from that of the host halo. Specifically, the concentration
parameter of the radial distribution of subhalos is fitted by 
\begin{equation}
c^*=c\left[1+\left(\frac{3m_s/M}{10^{-7}}\right)^{0.2}\right]^{-1}, 
\label{c_sub}
\end{equation}
where $c$ is the concentration parameter of the corresponding host halo.
Figure \ref{fig:cum_rad} shows the comparison of this fitting with the
model of \citet{oguri04c}. They agree reasonably well with each other
for a range of parameters we are interested in. 

\section{Linking Dark Halos and Subhalos to Galaxies}\label{sec:link}

We consider a central galaxy with luminosity $L$ which resides in a
dark halo with mass $M$. We adopt the mass-luminosity ($b_{\rm J}$
band) relation proposed by \citet{vale04}:
\begin{equation}
 L(M)=5.7\times 10^9h^{-2}L_{\odot}\frac{M_{11}^p}{\left[q+M_{11}^{(p-r)s}\right]^{1/s}},
\label{ml_halo}
\end{equation}
where $p=4$, $q=0.57$, $r=0.28$, $s=0.23$, and $M_{11}\equiv M/(10^{11}h^{-1}M_\odot)$.
Basically this relation asymptote to $L\propto M^4$ at low mass end and
$L\propto M^{0.28}$ at high mass end. \citet{vale04} claimed that the
same relation can be applied to dark halo substructures (with mass
$m_s$) and satellite galaxies (with luminosity $L_s$): 
\begin{equation}
 L_s(m_s)=5.7\times 10^9h^{-2}L_{\odot}\frac{m_{11}^p}{\left[q+m_{11}^{(p-r)s}\right]^{1/s}},
\label{ml_sub}
\end{equation}
where $m_{11}\equiv m/(10^{11}h^{-1}M_\odot)$ and $m$ is the mass of
subhalos before the tidal stripping of the outer part of
subhalos. We adopt the approximation relation $m=3m_s$, which was
also assumed in \citet{vale04}, throughout the paper for simplicity. 

While \citet{vale04} adopted monotonic one-to-one correspondence between
halo/subhalo masses and  resident galaxy luminosities,
\citet{cooray05a} showed that it is essential to assign log-normal
scatter to the mass-luminosity relations to recover the tail of the
observed luminosity function. Following them, we assume the log-normal
luminosity distribution of 
\begin{equation}
  p(L|M)=\frac{1}{\sqrt{2\pi}\ln(10)\Sigma
   L}\exp\left\{-\frac{\log_{10}\left[L/L(M)\right]^2}{2\Sigma^2}\right\}, 
\label{ml_scat}
\end{equation}
with $\Sigma=0.25$, and the same distribution with mean of $L_s(m_s)$ for $L_s$.  

Although this model is successful in explaining many observations such
as the occupation number, the luminosity function of clusters and the group
luminosity function, it cannot reproduce the strong dependence of the
faint end slope of the CLF on the halo mass: Observationally, it changes
from $\sim -1.2$ at the cluster scale to $\sim 0$ at galactic scales.
Since the mass function of subhalos at low mass is
$dn/dm_s\propto m_{\rm s}^{-\beta}$ where $\beta\sim 1.9$
(see \S \ref{sec:sub}), the mass-luminosity relation
$L_s(m_s)\propto m_s^\eta$ yields the CLF $\propto L_s^{-1-0.9/\eta}$.  
Therefore the above mass-luminosity relation implies the faint end slope of
$\sim -1.2$. It is also found that even if we modify the value of $\eta$ 
we cannot make the faint end slope shallower than $-1$. Hence to account
for the observed trend we need extra mechanism. Following \citet{cooray05b}, 
we introduce an efficient function to suppress the number of satellite
galaxies in galactic halos:
\begin{equation}
f(m_s)=0.5\left[1+{\rm erf}\left(\frac{\log m-\log m_c}{\sigma_m}\right)\right],
\label{eff}
\end{equation}
where $\sigma_m=1.5$ and the cutoff mass $m_c$ depends on the host halo
mass $M$:
\begin{equation}
m_c=10^{11}h^{-1}M_\odot\exp\left[-\left(\frac{M}{5\times10^{13}h^{-1}M_\odot}\right)^2\right].
\end{equation}
For less massive ($M\la 10^{13}h^{-1}M_\odot$) halos, the efficiency
function behaves as $f(m_s)\rightarrow 0$ when $m_s \ll
10^{11}h^{-1}M_\odot$ and $f(m_s)\rightarrow 1$ when $m_s \gg
10^{11}h^{-1}M_\odot$. At cluster scales $M\ga 10^{14}h^{-1}M_\odot$ the
efficiency function is almost always unity. 

We neglect the redshift evolution of the relations presented in this
section, since most lens galaxies are at $z<1$ where the evolution is
expected to be mild. 

\section{Computing the Image Separation Distribution}\label{sec:p}

To compute lensing cross sections, we have to convert galaxy luminosities
to any quantities which characterize the mass distribution of galaxies. To
do so, we use the scaling relations between the galaxy luminosity $L$,
velocity dispersion $\sigma$, and effective radius $R_0$. Specifically,
we adopt those derived from 9,000 early-type galaxies in the SDSS 
\citep{bernardi03}: 
\begin{equation}
 \frac{L}{10^{10.2}h_{70}^{-2}L_\odot}=\left(\frac{\sigma}{10^{2.197}{\rm rm\,s^{-1}}}\right)^{4.0},
\label{ltosig}
\end{equation}
\begin{equation}
 \frac{L}{10^{10.2}h_{70}^{-2}L_\odot}=\left(\frac{R_0}{10^{0.52}h_{70}^{-1}{\rm kpc}}\right)^{1.5},
\label{ltor0}
\end{equation}
where $L$ is $g$-band galaxy luminosity and we defined $h_{70}\equiv
h/0.7$. Note that the effective radius is related to the core radius of
Hernquist profile (eq. [\ref{hernquist}]) by $r_{\rm b}=0.551R_0$. For
simplicity, we neglect tiny difference of $b_J$-band and $g$-band
luminosities and regard $L$ in the above equations as the same
luminosities as those in \S \ref{sec:link}. Although the fundamental
plane can describe the correlation more tightly, we adopt these
relations between two observables because they are much easier to
handle. We also convert luminosities of galaxies to stellar masses
assuming the universal stellar mass-to-light ratio of
$\Upsilon=3.0h_{70}M_\odot/L_\odot$.  

One of the most important ingredient in computing lensing probabilities
is magnification bias which is determined by the source luminosity
function. Below we assume a power-law $\Phi(L)\propto
L^{-\xi}$, which makes the computation of magnification bias easier. 

\subsection{Dark Halo Component}\label{sec:p_halo}

As discussed in \S \ref{sec:dh}, we use the improved adiabatic
compression model of \citet{gnedin04} to compute the total mass
distribution of a galaxy plus dark halo and hence to derive its lensing
cross section. First, we derive the luminosity of a given halo with
mass $M$ using the mass-to-luminosity relation (\ref{ml_halo}), and
estimate the cooled baryon fraction $f_{\rm b}$ and the normalized
core radius $x_{\rm b}$ from the stellar mass-to-light ratio
$\Upsilon$ and equation (\ref{ltor0}), respectively. From the mass
distribution, we compute an image separation $\theta$ and biased cross
section $\sigma_{\rm lens}$: We use approximations proposed by
\citet{oguri02a} to compute these quantities. Then the image
separation distribution can be obtained by integrating the cross
section: 
\begin{equation}
\frac{dP_{\rm h}}{d\theta}=\int dz_l (1+z_l)^3\frac{cdt}{dz_l}\int dM
 \frac{dn}{dM}\sigma_{\rm lens}\delta(\theta-\theta(M)).
\label{prob_halo}
\end{equation}
The final probability distribution is obtained by averaging equation
(\ref{prob_halo}) over the probability distribution functions (PDFs) of
the concentration parameter and the mass-to-luminosity relation.

\subsection{Subhalo Component}

The projected mass distribution of a subhalo is approximated as
the sum of an SIS and external convergence $\kappa_{\rm ext}$ originates
from the mass associated with the host halo. We only consider the
external convergence and neglect the effect of the external shear, since it
has been shown that the external convergence is more effective to modify
the image separation distribution \citep{oguri05}. From the velocity
dispersion $\sigma$, the image separation $\theta$ and the biased cross
section $\sigma_{\rm lens}$ can be computed analytically \citep[see][for
the dependence on $\kappa_{\rm ext}$]{keeton04}:
\begin{equation}
\theta=\frac{2\theta_{\rm E}}{\left(1-\kappa_{\rm ext}\right)}=\frac{8\pi\left(\sigma/c\right)^2}{\left(1-\kappa_{\rm ext}\right)}\frac{D_{\rm ls}}{D_{\rm os}},
\label{the_sub}
\end{equation}
\begin{equation}
\sigma_{\rm lens}=\pi\left(D_{\rm ol}\theta_{\rm E}\right)^2\frac{2^\xi(1-\kappa_{\rm ext})^{-2(\xi-1)}}{3-\xi},
\end{equation}
where $D_{\rm os}$ is the angular diameter distance from the observer to
the source, etc. The external convergence is computed from the mass
distribution of the host halo after baryon cooling (\S \ref{sec:dh}). Since
the external convergence is written as a function of the projected
radius $R$, the PDF of the external convergence becomes 
\begin{equation}
p(\kappa_{\rm ext})d\kappa_{\rm ext}=f_R(R)\frac{d\kappa_{\rm ext}}{dR}dR,
\end{equation}
where $f_R(R)$ is the PDF of subhalos projected along
line-of-sight, which we adopt the projected NFW profile with its
concentration parameter given by equation (\ref{c_sub}):
\begin{equation}
f_R(R)dR=\frac{1}{h(c^*)}\int_{\phi_{\rm min}}^{\pi/2} d\phi\, g\left(\frac{c^* R/r_{\rm
 vir}}{\sin\phi}\right)\frac{c^*}{r_{\rm vir}}dR,
\end{equation}
\begin{equation}
\phi_{\rm min}=\sin^{-1}\left(\frac{R}{r_{\rm vir}}\right)
\end{equation}
\begin{equation}
g(x)=\frac{x}{(1+x)^2},
\end{equation}
\begin{eqnarray}
h(x)=\int_0^x dy \,g(y)&&\nonumber\\
&&\hspace*{-30mm}=
\left\{
\begin{array}{ll}
\frac{1}{1-x^2}\left[-1+\frac{2}{\sqrt{1-x^2}}{\rm
		arctanh}\sqrt{\frac{1-x}{1+x}}\right]
& (x>1)\\
\frac{1}{x^2-1}\left[1-\frac{2}{\sqrt{x^2-1}}{\rm
		arctan}\sqrt{\frac{x-1}{x+1}}\right]
& (x>1).
\end{array}
\right.
\end{eqnarray}
From these, we compute the image separation distribution as
\begin{eqnarray}
\frac{dP_{\rm s}}{d\theta}&=&\int dz_l (1+z_l)^3\frac{cdt}{dz_l}\int dM
 \frac{dn}{dM}\int dm_{\rm s}f(m_{\rm s})\frac{dN}{dm_{\rm s}}\nonumber\\
 &&\times \int d\kappa_{\rm ext}
  p(\kappa_{\rm ext}) \sigma_{\rm lens} \delta(\theta-\theta(\kappa_{\rm ext})).
\label{prob_sub}
\end{eqnarray}
As in \S \ref{sec:p_halo}, the final probability distribution is
obtained by averaging equation (\ref{prob_sub}) over the PDFs of
the concentration parameter of host halos (eq. [\ref{p_c}]) and the
mass-to-luminosity relation (eq. [\ref{ml_scat}]). We set the upper
limit of the integral over $\kappa_{\rm ext}$ 
to $0.9$, because in the events beyond the limit ($\kappa_{\rm
ext}\ga0.9$)  subhalos should fall inside the critical radius of the
host halo and therefore they should be included in lensing by dark halos
computed in \S \ref{sec:p_halo}.

We note that the inclusion of the efficient function (eq. [\ref{eff}])
implies that a part of low-mass subhalos remains dark, i.e., does not
harbor a galaxy. We neglect lensing by these dark subhalos because their
contribution to total lensing probability distributions is rather small 
unless the inner profile of subhalos is very steep \citep{ma03,li03}.

The total image separation distribution is simply given by a sum of
equations (\ref{prob_halo}) and (\ref{prob_sub}):
\begin{equation}
\frac{dP_{\rm t}}{d\theta}=\frac{dP_{\rm h}}{d\theta}+\frac{dP_{\rm s}}{d\theta}.
\label{prob_tot}
\end{equation}

\section{The Image Separation Distribution}\label{sec:sepdist}

In this section, we compute image separation distributions for halos
(central galaxies) and subhalos (satellite galaxies). In the specific
examples below, we fix the source redshift to $z_s=2.0$ which is typical
for lensed quasars. The magnification bias is computed assuming
a power-law luminosity function with $\xi=2.1$ which is the luminosity
function of the CLASS survey \citep{myers03}. First we will show the
image separation distributions for dark halos and subhalos
separately, and next we see the contribution of each component to the
total distribution. 

\subsection{Dark Halo Component}

\begin{figure}
\begin{center}
 \includegraphics[width=1.0\hsize]{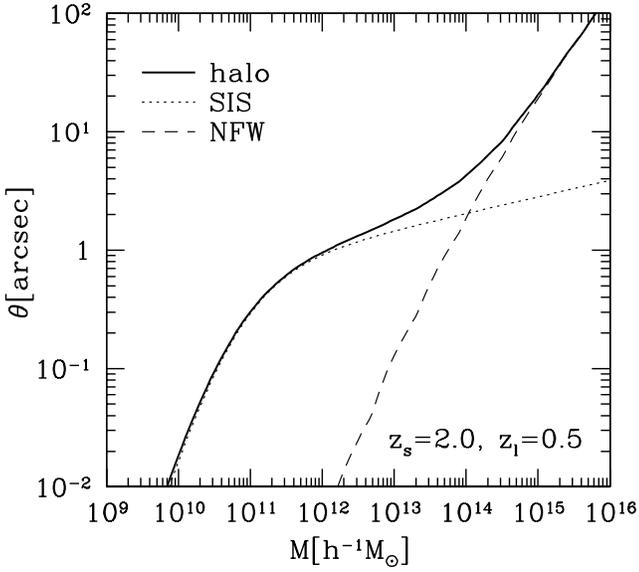}
\end{center}
\caption{The image separations of lensing by dark halos after baryon
 cooling are plotted as a function of halo masses. In this
 plot, we do not take the PDFs of the mass-to-luminosity relation and
 concentration parameter into account. The lens redshift is fixed to
 $z_l=0.5$. For comparison, we also show cases that we neglect baryon
 infall (NFW; {\it dashed line}) and we adopt an SIS approximation for
 the mass distribution of the halos after baryon infall (SIS; {\it
 dotted line}). Note that the velocity dispersion for SIS is derived
 from the scaling relation of equation (\ref{ltosig}).
\label{fig:sep}}
\end{figure}

Before going to the image separation distribution, we check the image
separation as a function of halo masses to see the effect of baryon
infall on lensing, which is shown in Figure \ref{fig:sep}. Two special
cases are also shown for reference: One is neglecting baryon infall and
leaving halos as NFW, and the other is adopting an SIS approximation
with the velocity dispersion inferred from the luminosity of the central
galaxy. It is found that the image separations agree well with those of
SIS at low-mass ends, which is consistent with observations
\citep[e.g.,][]{rusin03,rusin05}. With increasing masses, image
separations begin to deviate from SIS from $\theta\sim 1''$ 
($M\sim 10^{12}h^{-1}M_\odot$). At $\theta\ga 10''$ ($M\ga 3\times
10^{14}h^{-1}M_\odot$) image separations are quite similar to 
those of NFW. This figure justifies to some extent the two-population
model in which SIS and NFW lenses are considered, but we note that at
$M\sim 10^{13}-10^{14}h^{-1}M_\odot$ things are too complicated to be
described by such simple two-population model: Central galaxies are
mainly responsible for lensing, but halo masses outside the lensing
galaxies boost the image separation significantly, making the SIS
approximation inaccurate.

\begin{figure}
\begin{center}
 \includegraphics[width=1.0\hsize]{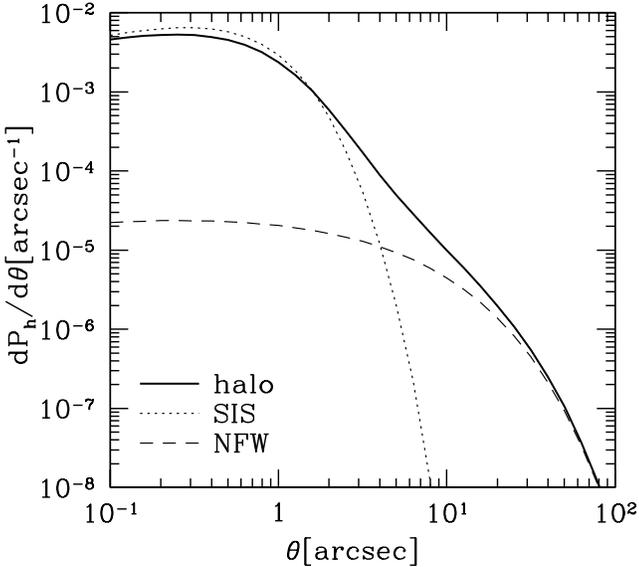}
\end{center}
\caption{The image separation distribution from the dark halo
 component (eq. [\ref{prob_halo}]). Lines are same as Figure
 \ref{fig:sep}. 
\label{fig:prob_h}}
\end{figure}

We show the image separation distribution in Figure \ref{fig:prob_h}.
As in Figure \ref{fig:sep}, at small- and large-separation regions
the probabilities are close to those of SIS and NFW, respectively, but
at $3''\la\theta\la10''$ it is difficult to describe lensing
probabilities as either SIS or NFW alone. The lensing probabilities
also differ from the sum of those of SIS and NFW. Therefore, we conclude
that the image separation distribution cannot be described by a simple
two-population model: We need to model the transition of halo
mass distributions from SIS to NFW to predict correctly the
distribution from small- to large-image separations. 

\begin{figure}
\begin{center}
 \includegraphics[width=1.0\hsize]{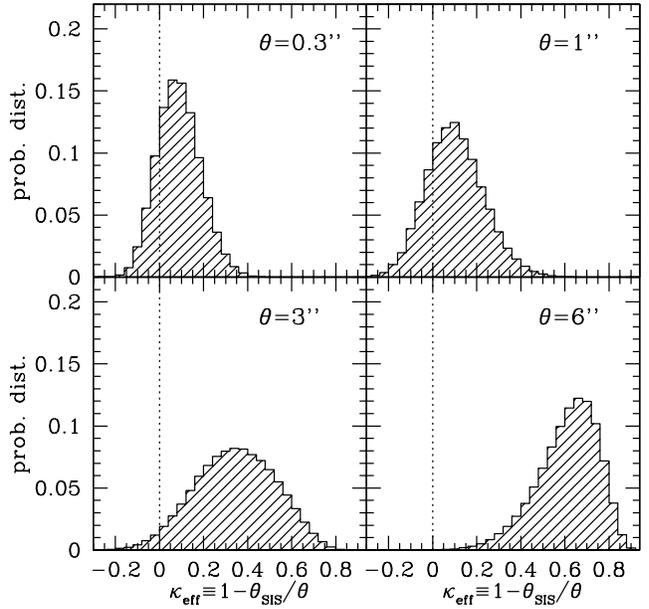}
\end{center}
\caption{The probability distribution of ``effective'' external
 convergence $\kappa_{\rm eff}$ (eq. [\ref{kap_eff}]) which is a
 measure of the deviation of the halo mass distribution from an SIS
 approximation of the corresponding central galaxy. We derive the
 distributions for several representative image separations:
 $\theta=0\farcs 3$ ({\it upper left}), $1''$ ({\it upper right}),
 $3''$ ({\it lower left}), and $6''$ ({\it lower right}). 
\label{fig:sep_h_dist}}
\end{figure}

We explore the transition from SIS to NFW more in terms of an 
``effective'' external convergence defined by
\begin{equation}
\kappa_{\rm eff}\equiv 1-\frac{\theta_{\rm SIS}}{\theta},
\label{kap_eff}
\end{equation}
where for each halo $\theta_{\rm SIS}$ is computed from the luminosity
of the central galaxy via equation (\ref{ltosig}). The case $\kappa_{\rm
eff}=0$ means that the Einstein radius of the halo is exactly the same
as that inferred from the luminosity of the central galaxy.
The definition is motivated by the description of the image separation
in the SIS plus external shear model (eq. [\ref{the_sub}]). We compute
the PDFs of $\kappa_{\rm eff}$ for several representative image
separations, which are shown in Figure \ref{fig:sep_h_dist}. As
expected, at $\theta\la 1''$ lens halos are on average close to an
SIS, being consistent with observations \citep[e.g.,][]{rusin03,rusin05}.
However, at $\theta\ga 3''$ they clearly have larger image separations
than those of corresponding SIS lenses. Therefore this Figure represents
another case for the deviation from simple SIS lenses at that image
separation region: At or beyond $\theta\ga 3''$ the image separations are
enhanced significantly by surrounding dark halos.

We also point out that $\kappa_{\rm ext}$ has rather broad distributions
even at small-separations $\theta\la 1''$, which comes from the scatters
of the concentration parameter (eq. [\ref{p_c}]) and the
mass-to-luminosity relation (eq. [\ref{ml_scat}])
around their medians. This scatter around SIS approximations may be
able to explain possible heterogeneous nature of lens galaxies
inferred from the combination of gravitational lensing and stellar
dynamics \citep{treu04} or time delay measurements between multiple
images of lensed quasars \citep{kochanek05}. 

\subsection{Subhalo Component}

\begin{figure}
\begin{center}
 \includegraphics[width=1.0\hsize]{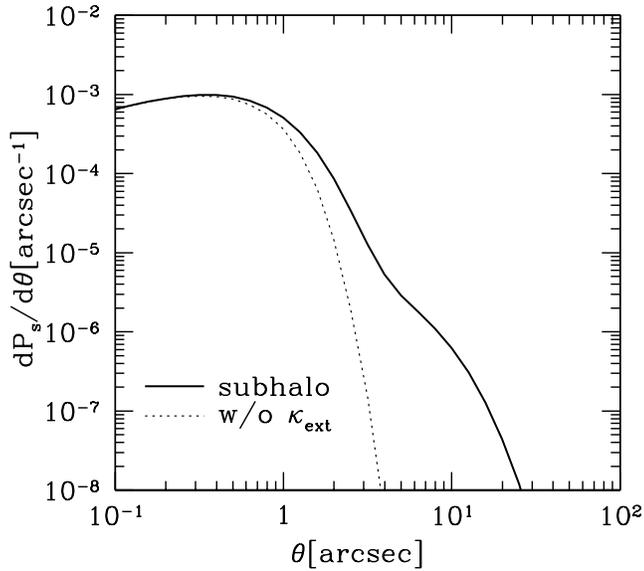}
\end{center}
\caption{The image separation distribution from the subhalo 
 component (eq. [\ref{prob_sub}]) is plotted by a solid line. 
We also calculate the distribution without including $\kappa_{\rm ext}$, 
 which is shown by a dotted line.
\label{fig:prob_s}}
\end{figure}

We show the image separation distribution from the subhalo component in
Figure \ref{fig:prob_s}. We find that the distribution
has a peak at around $\theta\sim 0.5''$, and rapidly decreases with
increasing $\theta$. It is also seen that the shape of the curve changes
at around $\theta\sim 5''$ over which the probability decreases less
rapidly. We attribute this behavior to the contribution of very large
external convergence ($\kappa_{\rm ext}\ga 0.5$), because the
probabilities are dominated by such extreme events (see below). This
means that the result is somewhat sensitive to the upper limit of 
$\kappa_{\rm ext}$ which we set $0.9$. However, as we will see in the
next subsection, this does not cause any significant problems since at
that image separation region the fraction of subhalo lensing in the
total lensing probabilities are rather minor.  

As shown in \citet{oguri05}, the external convergence affects the image
separation distribution significantly, particularly at the tail of the
distribution $\theta\ga 3''$. To show this explicitly, in Figure 
\ref{fig:prob_s} we also plot the distribution without including
$\kappa_{\rm ext}$.  It is found that the external convergence enhances
the lensing probabilities at $\theta\ga1''$ where the probabilities are
decreasing function of $\theta$. This is consistent with the finding by 
\citet{oguri05}. The enhancements are quite significant; factors of
$\sim 6$ at $\theta=2''$ and $\sim 80$ at $\theta=3''$. Note that these
numbers are much larger than the results of \citet{oguri05}, because
here we concentrate on satellite galaxies that should have larger
external convergence on average than central galaxies. Our result
indicates that it is very important to include the external convergence
of satellite galaxies, which originates from the host halo, to make
reliable predictions of the image separation distribution for the
subhalo component.

\begin{figure}
\begin{center}
 \includegraphics[width=1.0\hsize]{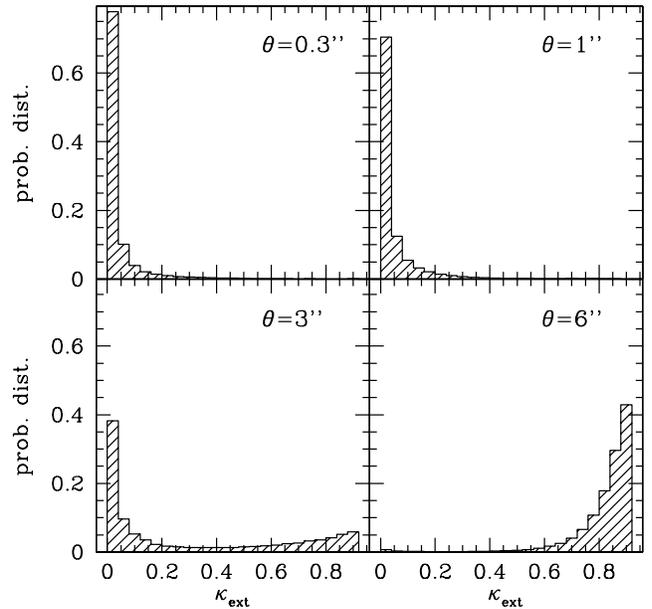}
\end{center}
\caption{The distribution of external convergence $\kappa_{\rm ext}$
 for the subhalo population. The PDFs for several different image
 separations are shown: $\theta=0\farcs 3$ ({\it upper left}), $1''$
 ({\it upper right}), $3''$ ({\it lower left}), and $6''$ ({\it lower
 right}). 
\label{fig:env}}
\end{figure}

It is interesting to compute the posterior PDF of $\kappa_{\rm ext}$ to
see expected environments of lensed satellite galaxies, as done by
\citet{oguri05}. In Figure \ref{fig:env}, we show the PDFs for
several different image separations. As expected, they are strong
function of the image separation: Larger-separation lenses tend to lie
in more dense environments. For instance, at $\theta\la 1''$ only 
$\sim 10\%$ of lensing by satellite galaxies have large external
convergence $\kappa_{\rm ext}>0.1$, while at $\theta\sim 3''$ more
than half of lenses are accompanied by such large external
convergence. Lenses with $\theta\sim 6''$ are dominated by those with
very large $\kappa_{\rm ext}$, which mean that lensing of such large
image separations occurs only when satellite galaxies lie close to the
center of their host halos on the projected two-dimensional lens
plane. 

\subsection{Total Distributions}\label{sec:total}

\begin{figure}
\begin{center}
 \includegraphics[width=1.0\hsize]{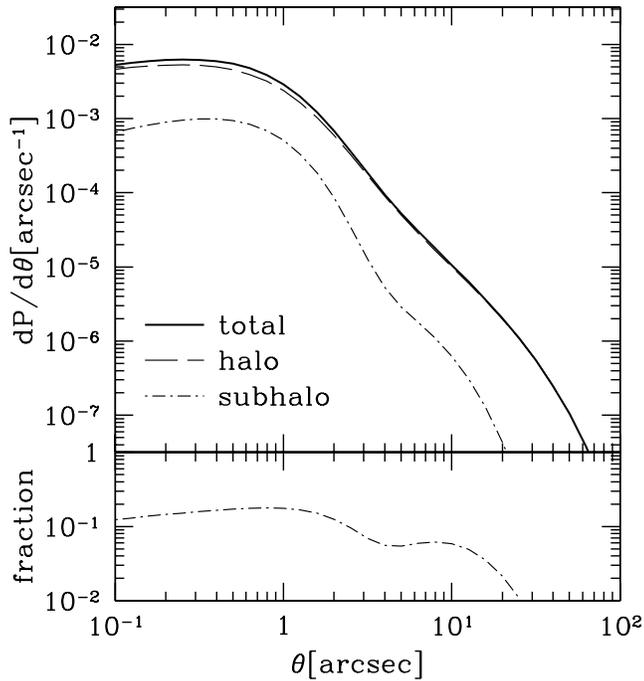}
\end{center}
\caption{{\it Upper:} The total image separation distribution $dP_{\rm
 t}/d\theta$ (eq. [\ref{prob_tot}]) is plotted by a solid line. The
 contributions of halo and subhalo components are also shown by dashed
 and dash-dotted lines, respectively. {\it  Lower:} The ratio of the
 distribution for the subhalo population to the total distribution,
 $(dP_{\rm s}/d\theta)/(dP_{\rm t}/d\theta)$, is plotted as a function
 of the image separation.
\label{fig:prob}}
\end{figure}

In this subsection, we put together the results in the previous
subsections to derive the total image separation of strong lenses
(eq. [\ref{prob_tot}]). 

Figure \ref{fig:prob} plots the total image separation distribution as 
well as distributions for the halo and subhalo populations. We find that 
the halo population dominates the lensing probability at all image 
separations. However, the contribution of the subhalo population is also 
significant: The fraction takes the maximum value of $\sim 0.2$ at around
$\theta=1''$, and decreases rapidly at $\theta\ga 3''$. Therefore, we 
conclude that the subhalo population should not be ignored for the 
accurate prediction of the image separation distribution.
Since most strong lenses have image separations of $\sim 1''-2''$, 
our model predicts that $10\%-20\%$ of lenses are produced by 
satellite galaxies and $80\%-90\%$ is caused by central galaxies. 

\begin{figure*}
\begin{center}
 \includegraphics[width=0.7\hsize]{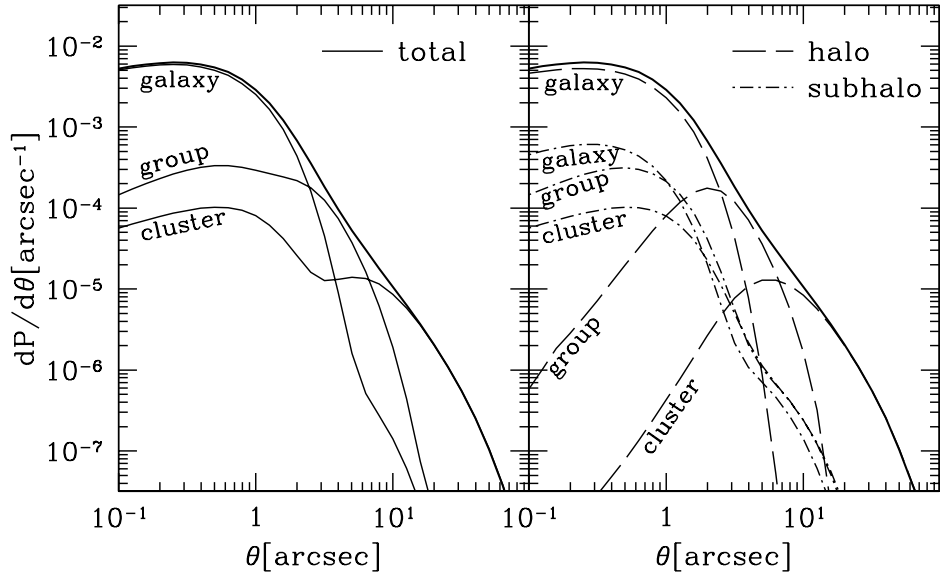}
\end{center}
\caption{The contributions of different types of halos on the image
  separation distribution. We consider the following three types:
  ``galaxy'' which is defined by (host-)halos with masses of 
 $M<10^{13}h^{-1}M_\odot$,  ``group'' by
  $10^{13}h^{-1}M_\odot<M<10^{14}h^{-1}M_\odot$, and ``cluster'' by
  $10^{14}h^{-1}M_\odot<M$. {\it Left:} The total 
  (halo plus subhalo populations) image separation for each type is
  plotted. The sum of distributions of three types is shown by thick
  solid line. {\it Right:} For each type, contributions of halo
  ({\it dashed}) and subhalo ({\it dash-dotted}) populations are
  shown separately. 
\label{fig:prob_dif}}
\end{figure*}

An advantage of our halo approach is the ability to obtain further
insight into the image separation distribution by computing the
contribution of the image separation distribution from different
mass intervals. For this purpose, we consider the following three
types of halos, ``galaxy'' (defined by (host-)halos with masses of
$M<10^{13}h^{-1}M_\odot$), ``group''
($10^{13}h^{-1}M_\odot<M<10^{14}h^{-1}M_\odot$) , and
``cluster'' ($10^{14}h^{-1}M_\odot<M$), and explore the contribution 
of each type. We show the result in Figure \ref{fig:prob_dif}.
We find that the distributions of group and cluster have tails toward
small-separations because of subhalo populations. Each type has quite
similar distributions of the subhalo population, while those of the
halo population are very different with each other. We also integrate
the distributions over the image separation to estimate the fraction
of each component to the total lensing probability, which is
summarized in Table \ref{table:frac}. Again, the fractions are very
different for halo populations, but the differences are much smaller
for subhalo populations. As a result, nearly half of lenses which lie in
groups and clusters are caused by satellite galaxies (subhalos), in
contrast with lenses in galactic halos which are dominated by the halo
population. Finally, we compute fractions of lenses in
lie in groups and clusters and find that they are $\sim 14\%$ and
$\sim 4\%$, respectively. These are roughly consistent with 
\citet{keeton00} who predicted that $\sim 20\%$ and $\sim 3\%$ of
lenses lie in groups and clusters, respectively (a part of the
differences may be ascribed to the different definitions of ``group''
and ``cluster''). 
\begin{table}
 \caption{The fraction of lensing probabilities, integrated over the
 image separation. The entry ``total'' indicates a sum of lensing
 probabilities for halo and subhalo populations. \label{table:frac}}
 \begin{tabular}{@{}cccc}
  \hline
   type & galaxy &  group & cluster \\
   ([$h^{-1}M_\odot$]) & ($<10^{13}$) & ($10^{13}-10^{14}$) &
   ($>10^{14}$) \\
  \hline
  halo & $0.74$ & $0.08$ & $0.02$ \\ 
  subhalo & $0.08$ & $0.06$ & $0.02$ \\
  total & $0.82$ & $0.14$ & $0.04$ \\
  \hline
 \end{tabular}
\end{table}

\begin{figure*}
\begin{center}
 \includegraphics[width=0.7\hsize]{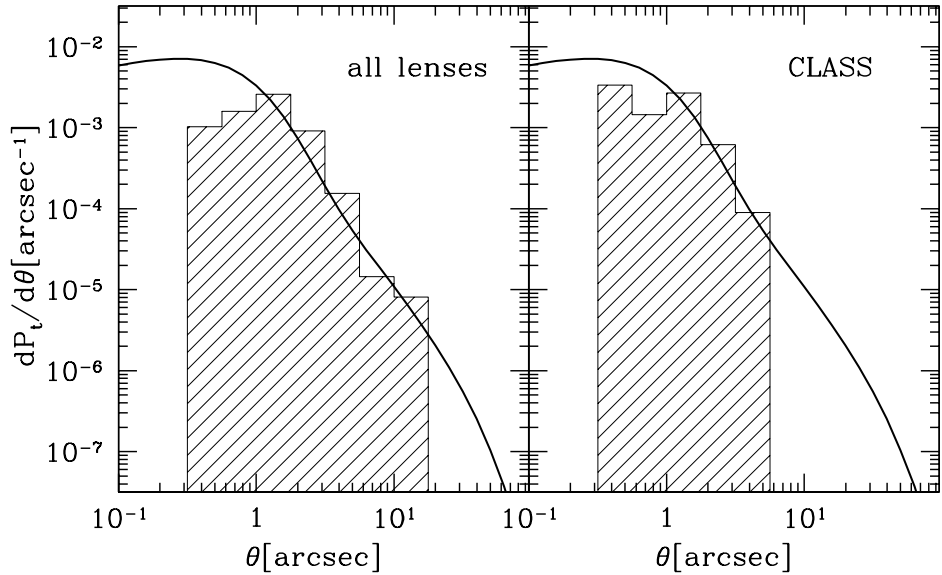}
\end{center}
\caption{Observed image separation distributions ({\it histograms})
 are compared with the theoretical prediction presented in Figure
 \ref{fig:prob}. We consider two observed distribution: (i) The
 distribution of all 75 lensed quasars discovered to date, and (ii)
 that of 22 lenses discovered in the CLASS \citep{myers03}. The
 observed distributions were shifted vertically to match their
 normalizations to the theoretical one. 
\label{fig:prob_obs}}
\end{figure*}

How well does the total image separation distribution derived here
account for observed image separation distributions? We compare our
model prediction with the following two observed distributions:

\begin{itemize}
\item All 75 lensed quasars discovered to date. This sample includes both 
optical and radio lenses and are quite heterogeneous, but it has an 
advantage of the large number of lenses including several 
intermediate-separation lenses ($3''\la \theta\la 7''$) as well as 
one large-separation lens whose separation is $\theta\sim15''$. 
\item 22 radio lenses discovered in the CLASS \citep{myers03}. Although 
this sample contains smaller number of lenses (and all lenses have 
$\theta<5''$), it is much more homogeneous and has well-defined selection 
functions. Although among the 22 lenses 13 were selected for 
a statistically well-defined subsample \citep{browne03}, we use all 22
lenses simply to increase statistics.
\end{itemize}

The results are shown in Figure \ref{fig:prob_obs}. We find that our model 
explains both observed distributions reasonably well. An exception is that 
for all lenses the observed number of sub-arcsecond ($\theta< 1''$) 
lenses appears to be smaller than expected. However, this is clearly due to 
a selection effect that sub-arcsecond lenses are difficult to find 
particularly in optical lens surveys. Indeed, for the CLASS lenses, 
which is a lens survey in a radio band, the discrepancy is less significant.
Thus we conclude that our model well accounts for the observed image 
separation distributions.

\begin{figure*}
\begin{center}
 \includegraphics[width=0.7\hsize]{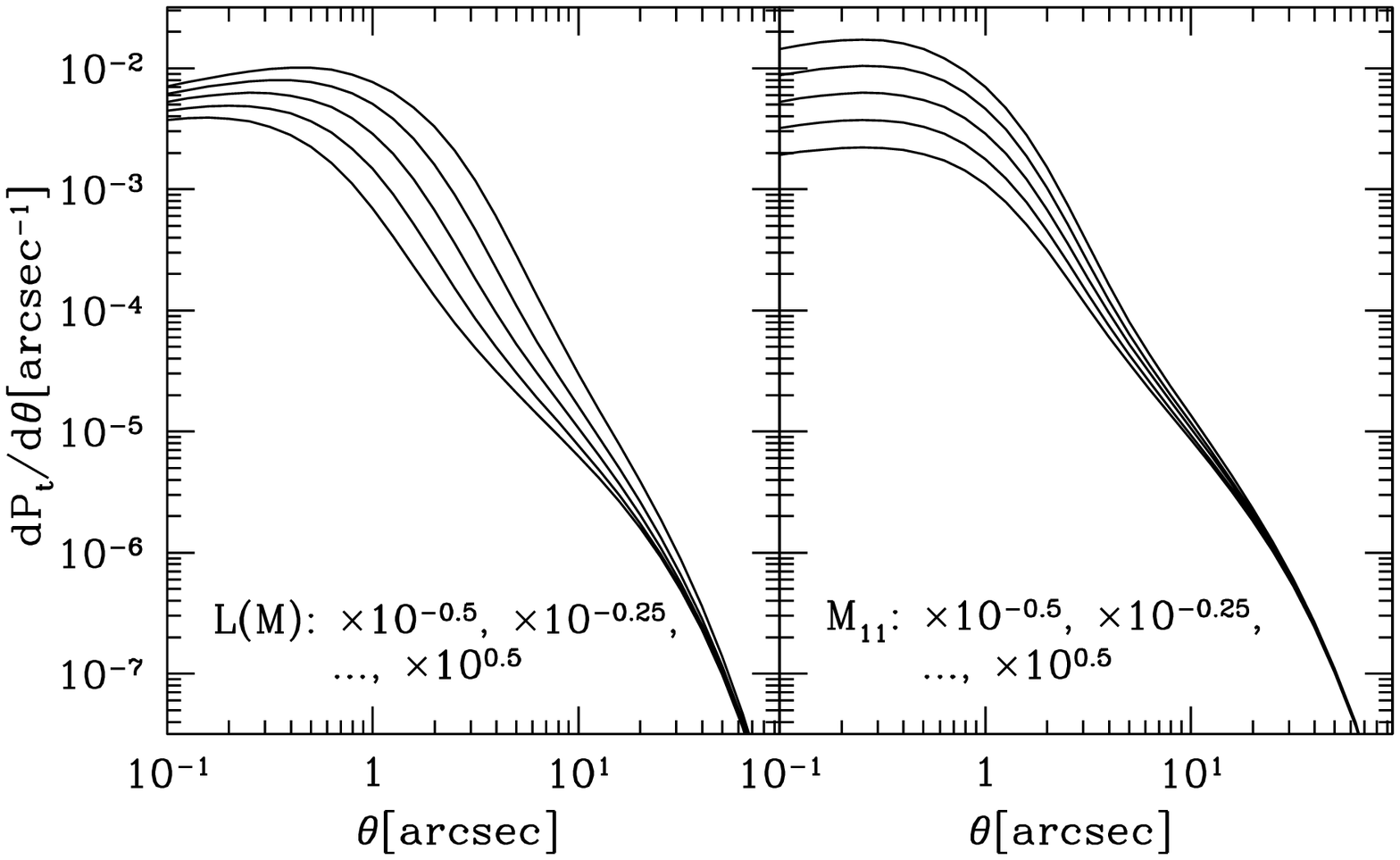}
\end{center}
\caption{The dependence of the total image separation distribution on
the mass-to-luminosity relations $L(M)$ (eq. [\ref{ml_halo}]) and 
$L_s(M_s)$ (eq. [\ref{ml_sub}]). {\it Left:} We change the overall 
normalizations of $L(M)$ and $L_s(M_s)$ by multiplying numerical 
factors to them. From upper to lower lines, we multiply $10^{0.5}$, 
$10^{0.25}$, $1$, $10^{-0.25}$, and $10^{-0.5}$. 
{\it Right:} We shift the characteristic mass scale $M_{11}$ and
$m_{11}$ in the mass-to-luminosity relations by multiplying the 
same factors to both $M_{11}$ and $m_{11}$ (upper lines correspond to
the cases that smaller factors are multiplied).
\label{fig:prob_ml}}
\end{figure*}

It should be noted that the theoretical predictions is based on many 
assumptions. The most important ingredient of our model is the 
mass-to-luminosity relation defined in equations (\ref{ml_halo})
and (\ref{ml_sub}). To see how the result is dependent on the adopted
relation, in Figure \ref{fig:prob_ml} we show the dependence of the 
total image separation distribution on the mass-to-luminosity relations.
As seen in the Figure, the shape (rather than the overall normalization) 
of the image separation distribution is quite sensitive to the 
mass-to-luminosity relation. This indicates that precise measurements
of the image separation distribution offer quite useful probe of 
the mass-to-luminosity relation.

\section{Summary and Discussions}\label{sec:sum}

In this paper, we have constructed a model of the image separation 
distribution. Our model is based on dark halos and subhalos which are 
linked to central and satellite galaxies respectively via simple 
mass-to-luminosity relation. For dark halos and central galaxies, we 
have considered baryon cooling in a dark matter halo using an improved 
adiabatic contraction model of \citet{gnedin04}. For satellite galaxies,
we have considered the mass associate with their host halo.
Our primary interest is to quantify the contribution of the subhalo
population which has been ignored in all previous studies.  Our results 
are summarized as follows:

\begin{itemize}
\item We predict that $80\%-90\%$ of lenses should be caused by central 
galaxies (halos) and $10\%-20\%$ of lenses should be produced by satellite
galaxies (subhalos). The fraction of the subhalo population takes the maximum
at $\theta\sim 1''$, and becomes smaller with increasing image 
separations.
\item The mass distributions of lensing halos at small ($\theta\la 3''$)
and large ($\theta\ga 10''$) image separation regions are close to those of
SIS and NFW, respectively. At $3''\la\theta\la10''$ the mass distributions 
are complicated and cannot be described by either SIS or NFW. 
\item For both halo and subhalo populations, already at $\theta\sim 3''$ the
effect of dark halos becomes significant, making a simple SIS
approximation worse. This means applying isothermal profiles to these
lenses results in biased estimates of parameters such as velocity
dispersions and the Hubble constant from time delays.
\item Our model predicts that lensing halos of small separation lenses are 
heterogeneous: Halo-by-halo differences of the shape of rotation curves and
the fraction of dark matter at image positions are large. 
This is a natural consequence of the scatters of the concentration parameter 
and the mass-to-luminosity relation around their medians.
\item In computing the image separation distribution for the subhalo component,
it is important to take account of the external mass which comes from
their host halos. 
This also implies very strong dependence of environments of lensing satellite 
galaxies on image separation distributions.
\item Halo and subhalo populations have quite different contributions of
  image separation distributions from different (host-)halo mass
  intervals. We predict that $\sim 14\%$ of lenses lie in groups, and
  $\sim 4\%$ in clusters. Almost half of lenses in groups and clusters
  are produced by satellite galaxies (subhalos), rather than central 
  galaxies (halos).
\item We have compared our model predictions with observed image separation 
distributions and found that they are in reasonable agreements with each
other. Since the shape of the image  separation distribution is rather sensitive
to the mass-to-luminosity relation, precise measurements of the image separation
distribution in well-defined statistical lens samples offer powerful probe
of the connection between dark halos and galaxies.
\end{itemize}

In summary, we have constructed a realistic model which predicts 
lensing probabilities from small to large image separations. The model
is quite useful in understanding lens populations as a function of the
image separation. The fraction of lensing by the subhalo population, 
which has been derived for the first time in this paper, is important
for not only an accurate prediction of the image separation distribution 
but also interpreting results of mass modeling of individual lens systems
since central and satellite galaxies may show rather different lensing
characteristics. 

We note that there are several ways to improve our modeling. First, we
have considered early-type galaxies only. Although this can be
justified because most of lensing galaxies are early-type, the very
existence of lensing by late-type galaxies indicates that the correct
model needs to take both the two galaxy  types into account. Since the
lensing by spiral galaxies has quite small image separation
$\theta\la1''$ and also is inefficient \citep{keeton98}, we expect
that the inclusion of late-type galaxies decreases the number of
sub-arcsecond lenses caused by central galaxies. In
addition, we have neglected the scatters of the relations  between
galaxy luminosities, velocity dispersions, and effective radii. The
scatters could affect the quantitative results, as the scatter in
the mass-to-luminosity relation is important.  The redshift
evolutions of relations are also neglected. However, if we assume the
evolution of early type galaxies at $z\la1$ is purely passive, it
implies that the mass distribution of lensing halos does not change
across the redshift. Therefore we
expect the effect of redshift evolution is rather small. More
importantly, we have assumed simple spherical halos. In reality, dark
halos are quite triaxial \citep{jing02} and the triaxiality enhances
lensing probabilities by a few factors at large-separations
\citep{oguri03,oguri04b,hennawi05a,hennawi05b}. On the other hand,
small-separation lensing probabilities are hardly affected by the
ellipticity of lens galaxies: \citet{huterer05} showed that the change
is less than a few percents unless quasars are very
bright. Large-separation lensing probabilities are also sensitive to
the inner slope of dark matter halos
\citep[e.g.,][]{keeton01b,wyithe01,takahashi01,li02,oguri02b}.
We not, however, that the mass distribution after baryon cooling is
rather insensitive to the inner slope of dark halos, thus the fraction
of lensing by satellite galaxies does not depend on the inner slope
very much. Refining our model by incorporating these is beyond the
scope of this paper, but is of interest for future studies.

\section*{Acknowledgments}
The author would like to thank C. Keeton, N. Dalal, J. Ostriker,
J. Lee, and A. Vale for discussions. 
The author is supported by JSPS through JSPS Research Fellowship for
Young Scientists.  


\appendix

\label{lastpage}

\end{document}